\documentclass{article}

\usepackage{amsmath}
\usepackage{amssymb}     
\usepackage{latexsym}       

\usepackage{epsfig}

\newcommand{\dda}{\mathord{\mbox{\makebox[0pt][l]{\raisebox{-.4ex}
                           {$\downarrow$}}$\downarrow$}}}
\newcommand{\dua}{\mathord{\mbox{\makebox[0pt][l]{\raisebox{.4ex}
                           {$\uparrow$}}$\uparrow$}}}
 
\newcommand{\bq}{\begin{quote}}
\newcommand{\eq}{\end{quote}}

\newcommand{\dom}{\mathrm{dom}}

\newcommand{\reals}{{\mathbb R}}
\newcommand{\IR}{{\mathbf I}\,\!{\mathbb R}}
\newcommand{\BX}{{\mathbf B}X}
\newcommand{\UX}{{\mathbf U}\,\!{\mathit X}}

\newcommand{\nat}{{\mathbb N}}
\newcommand{\rat}{{\mathbb Q}}
\newcommand{\Pnat}{{\mathcal P}\omega}

\newtheorem{Th}{Theorem}[section]
\newtheorem{theorem}[Th]{Theorem}
\newtheorem{proposition}[Th]{Proposition}     
\newtheorem{lemma}[Th]{Lemma}

\newtheorem{definition}[Th]{Definition} 
\newtheorem{example}[Th]{Example}

\begin{document}
\date{}
\title{P{\large ARTIALITY IN PHYSICS}}

\author{
B{\footnotesize OB} C{\footnotesize OECKE} and 
K{\footnotesize EYE} M{\footnotesize ARTIN}\\$ $\\
{\small 
Oxford University Computing Laboratory,
}\\
{\small 
Wolfson Building, Parks Road, Oxford OX1 3QD.
}\\
{\small 
\texttt{Bob.Coecke@comlab.ox.ac.uk\ $\cdot$\ Keye.Martin@comlab.ox.ac.uk}
}
}

\maketitle

\vspace{-8mm}

\begin{abstract}
We revisit the standard axioms of domain theory
with emphasis on their relation to the concept
of partiality, explain how this idea arises
naturally in probability theory and quantum mechanics, and then
search for a mathematical setting capable
of providing a satisfactory unification of the two.
\end{abstract}

\section{Introduction}

Dana Scott\cite{scott:lec} introduced domains more than thirty years
ago as an appropriate mathematical universe for the
semantics of programming languages. A domain is a
partially ordered set with intrinsic notions
of completeness and approximation. Recently\cite{meandbob},
the authors have proven the existence of a natural domain theoretic
structure in probability theory and {quantum mechanics}. The way to understand 
this structure is with
the aid of the concept \em partiality. \em

To illustrate, in the domain $(\mathcal{P}\omega,\subseteq)$,
the powerset of the natural numbers ordered by inclusion, a finite
set will be partial, while the set $\omega$ will be total.
In the domain $(\Sigma^\infty,\sqsubseteq)$, the domain of bit streams with
the prefix order,
a finite string is partial, the infinite strings are total.
In the domain $(\IR,\sqsubseteq)$, the collection of compact intervals $[a,b]$
of the real line ordered by reverse inclusion, 
an interval like $[p,q]$ with $p<q$ rational is
partial, while a one point interval $[x]$ representing
a real number is total. In the domain $(\Omega^n,\sqsubseteq)$, 
the $n$ dimensional mixed states in the spectral order (to be defined later),
a pure state is total, while mixed states which are not pure
are partial. In all the cases above,
total elements coincide with elements
which are \em maximal \em in the given order.

As we can see, the partiality idea arises naturally
in both computer science and physics. The idea
is important in computer science. We will
review results herein which show that 
by reasoning about density operators 
as partial and total objects, one can
derive the classical and quantum logics
of Birkhoff and von Neumann as special
order theoretic subsets. Because of this,
we conclude that partiality is also an important
idea in physics. Given that
the idea is important, the main question
asked in this paper is ``What is an appropriate
mathematical setting for discussing partiality?''

First we review the traditional axioms for domains,
which succeed at capturing the notion partiality
for objects like sets, strings and intervals. Then
we consider the Bayesian and spectral orders on
classical and quantum states, which are `domains'
that possess the same
notion of completeness as do classical domains, 
but differ in that they offer a new notion of approximation.
We then 
enumerate much of what we know about 
the order theoretic structure
of these two
new domains in the hope that it will
point the way for an inspired reader
to discover a proper generalization of classical domains
that will have desirable properties useful
to both physicists and computer scientists.

\section{Domain theory}

As we mentioned in the introduction, a domain
$(D,\sqsubseteq)$ is a partially ordered set with notions
of completeness and approximation. The completeness
can be used for example to prove fixed point theorems,
which themselves might be used to provide a semantics
for recursion, or establish the existence
of solutions to ordinary differential equations.
Explaining approximation is more difficult.

The order on a domain can be used to define many topologies,
some of which can be used to recover notions of limit
that we are familiar with from analysis. One use for
\em approximation \em -- which itself is a
relation $\ll$ contained in $\sqsubseteq$ -- is that it can clarify these topologies for us,
helping us to connect them to more familiar ideas. A more
subtle use for approximation is in formalizing the notion \em partiality. \em
To give a simple example, an object $x\in\mathcal{P}\omega$ approximates
something -- formally, there is $y$ with $x\ll y$ --
if and only if $x$ is finite.

\subsection{Order}

A \em poset \em is a partially ordered set, i.e., a set together
with a reflexive, antisymmetric and transitive relation.

\begin{definition}\em Let $(P,\sqsubseteq)$ be a partially ordered set.
  A nonempty subset $S\subseteq P$ is \em directed \em if $(\forall
  x,y\in S)(\exists z\in S)\:x,y\sqsubseteq z$. The \em supremum \em
  of $S\subseteq P$ is the least of all its upper bounds
  provided it exists. This is written $\bigsqcup S$.
\end{definition}

\begin{definition}\em For a subset $X$ of a poset $P$, set
\[\uparrow\!\!X:=\{y\in P:(\exists x\in X)\,x\sqsubseteq y\}\ \ \&\
\downarrow\!\!X:=\{y\in P:(\exists x\in X)\,y\sqsubseteq x\}.\] We
write $\uparrow\!x=\,\uparrow\!\{x\}$ and
$\downarrow\!x=\,\downarrow\!\{x\}$ for elements $x\in X$.
\end{definition}

A partial order allows for the derivation of
several intrinsically defined topologies. 

\begin{definition}\em  A subset $U$ of a poset $P$ is \em Scott open \em if
\begin{enumerate}
\item[(i)] $U$ is an upper set: $x\in U\ \&\ x\sqsubseteq y\Rightarrow
  y\in U$, and \item[(ii)] $U$ is inaccessible by directed suprema:
  For every directed $S\subseteq P$ with a supremum,
\[\bigsqcup S\in U\Rightarrow S\cap U\neq\emptyset.\]
\end{enumerate}
The collection of all Scott open sets on $P$ is called the \em Scott
topology. \em
\end{definition}

Unless explicitly stated otherwise, all topological statements about
posets are made with respect to the Scott topology.

\begin{proposition}
  A function $f:P\rightarrow Q$ between posets is continuous iff
\begin{enumerate}
\em\item[(i)]\em f is monotone: $x\sqsubseteq y\Rightarrow f(x)\sqsubseteq
  f(y).$ 
\em\item[(ii)]\em f preserves directed suprema: For every directed
  $S\subseteq P$ with a supremum, its image $f(S)$ has a supremum, and
\[f(\bigsqcup S)=\bigsqcup f(S).\]
\end{enumerate}
\end{proposition}

The \em completeness \em of domains comes
from the fact that directed sets have suprema:

\begin{definition}\em
A \em dcpo \em is a poset in which every directed subset has a
  supremum. The \em least element \em in a poset,
  when it exists, is the unique element $\bot$ with $\bot\sqsubseteq x$
for all $x$.
\end{definition}

Here is the most well-known
fixed point theorem in domain theory.

\begin{theorem} Let $f:D\rightarrow D$ be a Scott continuous map on a dcpo with
a least element. Then
\[\mathrm{fix}(f):=\bigsqcup_{n\geq 0}f^n(\bot)\]
is the least fixed point of $f$.
\end{theorem}
The set of \em maximal elements \em in a dcpo $D$ is
\[\max(D):=\{x\in D :\ \uparrow\!\!x=\{x\}\}.\]
Each element in a dcpo has a maximal
element above it.
\begin{example}\em Let $X$ be a compact Hausdorff space. Its \em upper space
  \em
\[\UX=\{\emptyset\neq K\subseteq X:K\mbox{ is compact}\}\]
ordered under reverse inclusion
\[A\sqsubseteq B\Leftrightarrow B\subseteq A\]
is a dcpo: For directed $S\subseteq\UX$, $\bigsqcup S=\bigcap S.$ 
A continuous map $f:X\rightarrow X$ induces a Scott continuous map
\[\bar{f}:\UX\rightarrow\UX::K\mapsto f(K)\]
and since $\bot=X\in\UX$, the fixed point theorem guarantees that
\[\mathrm{fix}(\bar{f}):=\bigcap f^n(X)\]
is the least fixed point of $\bar{f}$. That is, $f$ has a unique
largest invariant set $K=f(K)$. If $f$ were a contraction,
then we would have $K=\{x^*\}$, where $x^*$ is the 
unique fixed point of $f$.

It is interesting here that the space $X$ can be recovered
from $\UX$ in a purely order theoretic manner: It can be shown
that
\[X\simeq\max(\UX)=\{\{x\}:x\in X\}\]
where $\max(\UX)$ carries the relative Scott topology it 
inherits as a subset of $\UX.$ Several constructions
of this type are known, especially for Hilbert spaces.
This illustrates one way that an order can implicitly describe a topology.
\end{example}

\subsection{Approximation and continuity}

Domains are posets that carry intrinsic notions of approximation
and completeness.

\begin{definition}\em
  For elements $x,y$ of a poset, write $x\ll y$ iff for all directed
  sets $S$ with a supremum,
\[y\sqsubseteq\bigsqcup S\Rightarrow (\exists s\in S)\:x\sqsubseteq s.\]
We set $\dda x=\{a\in D:a\ll x\}$ and $\dua x=\{a\in D:x\ll a\}$.
\end{definition}
For the symbol ``$\ll$,'' read ``approximates.'' 

\begin{definition}\em
  A \em basis \em for a poset $D$ is a subset $B$ such that $B\cap\dda x$
  contains a directed set with supremum $x$ for all $x\in D$.  A poset is
  \em continuous \em if it has a basis. A poset is $\omega$-\em continuous \em
  if it has a countable basis.
\end{definition}
A \em continuous dcpo \em is a continuous poset which is also a dcpo.
\begin{example}\em
  The collection of functions
\[\Sigma^\infty=\{\:s\:|\:s:\{1,\dots,n\}\rightarrow\{0,1\},
0\leq n\leq\infty\:\}\] ordered by extension
\[s\sqsubseteq t\Leftrightarrow|s|\leq|t|\ \&\ (\:\forall\:1\leq
i\leq|s|\:)\:s(i)=t(i),\] where $|s|$ is the cardinality
of $\dom(s)$, is an $\omega$-algebraic dcpo:
\begin{itemize}
\item For directed $S\subseteq\Sigma^\infty,$ $\bigsqcup S=\bigcup S$,
\item $s\ll t\Leftrightarrow s\sqsubseteq t\ \&\ |s|<\infty,$
\item $\{s\in\Sigma^\infty:|s|<\infty\}$ is a countable basis for
  $\Sigma^\infty,$ \item The least element $\bot$ is the unique $s$
  with $|s|=0.$ 
\end{itemize}
\end{example}
The next example is due to Scott\cite{scott:lec}.
\begin{example}\em  The collection of compact intervals of the real line
\[\IR=\{[a,b]:a,b\in\reals\ \&\ a\leq b\}\]
ordered under reverse inclusion
\[[a,b]\sqsubseteq[c,d]\Leftrightarrow[c,d]\subseteq[a,b]\]
is an $\omega$-continuous dcpo:
\begin{itemize}
\item For directed $S\subseteq\IR$, $\bigsqcup S=\bigcap S$, \item
  $I\ll J\Leftrightarrow J\subseteq\mbox{int}(I)$, and \item
  $\{[p,q]:p,q\in\rat\ \&\ p\leq q\}$ is a countable basis for $\IR$.
\end{itemize}
The domain $\IR$ is called the \em interval domain.\em
\end{example}
Approximation can help explain
the Scott topology on a continuous dcpo.

\begin{theorem}
  The collection $\{\dua x:x\in D\}$ is a basis for the Scott topology
  on a continuous dcpo.
\end{theorem}
The last result also holds for continuous posets.

\begin{example}\em
  A basic open set in $\IR$ is
\[\dua[a,b]=\{x\in\IR:x\subseteq(a,b)\}\]
while a basic open set in $\Sigma^\infty$ is
\[\uparrow\!\!s=\{t\in\Sigma^\infty:(\exists u\in\Sigma^\infty)\,t=s u\}\]
for $s$ finite.
\end{example}
With the algebraic domains, we
come closest to the ideal of `finite approximation.'

\begin{definition}\em
  An element $x$ of a poset is \em compact \em if $x\ll x$. A poset is \em
  algebraic \em if its compact elements form a basis; it is
  $\omega$-\em algebraic \em if it has a countable basis of compact
  elements.
\end{definition}

\begin{example}\em  The powerset of the naturals
\[\mathcal{P}\omega=\{x:x\subseteq\omega\}\] 
ordered by inclusion
\[x\sqsubseteq y\Leftrightarrow x\subseteq y\] 
is an $\omega$-algebraic dcpo:
\begin{itemize}
\item For directed set $S\subseteq\mathcal{P}\omega$,
  $\bigsqcup S=\bigcup S$,
\item $x\ll y\Leftrightarrow x\sqsubseteq y\ \&\ x\ \mbox{is
    finite},$ and
\item $\{x\in\mathcal{P}\omega:x\ \mbox{is finite}\}$ is a
  countable basis for $\mathcal{P}\omega.$
\end{itemize}
\end{example}
The next domain is of central importance in recursion theory
(Odifreddi\cite{odifreddi:recursion}).
\begin{example}\em  The set of partial mappings on the naturals
\[[\nat\rightharpoonup\nat]=\{\:f\:|\:f:\nat\rightharpoonup\nat\mbox{
is a partial map}\}\] ordered by extension
\[f\sqsubseteq g\Leftrightarrow \dom(f)\subseteq\dom(g)\ \&\
f=g\ \mbox{on }\dom(f)\] is an $\omega$-algebraic dcpo:
\begin{itemize}
\item For directed set $S\subseteq[\nat\rightharpoonup\nat]$,
  $\bigsqcup S=\bigcup S$,
\item $f\ll g\Leftrightarrow f\sqsubseteq g\ \&\ \dom(f)\ \mbox{is
    finite},$ and
\item $\{f\in[\nat\rightharpoonup\nat]:\dom(f)\mbox{ finite}\}$ is a
  countable basis for $[\nat\rightharpoonup\nat].$
\end{itemize}
\end{example}

\subsection{Measurement}

A few of the ideas
that the study of measurement\cite{martin:thesis}
has led to include an informatic derivative, new fixed 
point theorems, the derivation of distance from content, 
techniques for treating continuous and discrete 
processes and data in a unified manner, a `first order' view 
of recursion based on solving 
renee equations $\varphi=\delta+\varphi\circ r$ uniquely which 
establishes surprising connections between order and computability,
and various approaches to complexity. 

The original idea was that if a domain gave a formal
account of `information,' then a measurement on a domain
should give a formal account of `information content.' 
There is a stark difference between the view of
information content taken in the study of measurement,
and utterances of this phrase made elsewhere; it is this: 
Information content is a structural relationship between two classes
of objects which, generally speaking, arises
when one class may be viewed as a simplification of the other.
The process by which a member of one class 
is simplified and thereby `reduced' to an element
of the the other is what we mean by `the measurement process' in domain
theory\cite{martin:measure}.

One of the classes may well be a subset of real numbers, but  
the `structural relationship' underlying content should not be forgotten.
For example, this principle can be
taken as the basis for a new approach to the study of entanglement.

\begin{definition}\em A Scott continuous map $\mu:D\rightarrow E$ 
between dcpo's is said to \em measure the content of $x\in D$ 
\em if 
\[x\in U\Rightarrow(\exists\varepsilon\in 
\sigma_E)\,x\in\mu_\varepsilon(x)\subseteq U,\] 
whenever $U\in\sigma_D$ is Scott open and 
\[\mu_\varepsilon(x):=\mu^{-1}(\varepsilon)\,\cap\downarrow\!\!x\] 
are the elements $\varepsilon$ close to $x$ in content. The map 
$\mu$ \em measures \em $X$ if it measures the content of each $x\in X$. 
\end{definition} 

\begin{definition}\em A \em measurement \em is a Scott continuous 
map $\mu:D\rightarrow E$ between dcpo's that measures 
${\ker\mu:=\{x\in D:\mu x\in\max(E)\}}$. 
\end{definition} 

The case $E=[0,\infty)^*$ is especially important. Then 
$\mu$ is a measurement iff for all $x\in D$ with $\mu x=0$, 
\[x\in U\Rightarrow(\exists\varepsilon>0)\,x\in\mu_\varepsilon(x)\subseteq
U,\] 
whenever $U\subseteq D$ is Scott open. The elements $\varepsilon$ close to
$x\in\ker\mu$ are 
then given by 
\[\mu_\varepsilon(x):=\{y\in D:y\sqsubseteq x\ \&\ |\mu x-\mu 
y|<\varepsilon\},\]
where for a \em number \em $\varepsilon>0$ and $x\in\ker\mu$, 
we write $\mu_\varepsilon(x)$ for $\mu_{[0,\varepsilon)}(x).$
In this case, $\mu x$ measures the \em uncertainty \em in 
$x$. Thus, an object with measure zero ought to have no uncertainty, 
which means it should be maximal. 

\begin{lemma} If $\mu$ is a measurement, then 
$\ker\mu\subseteq\max(D)$. 
\end{lemma} 

In fact, measurements are \em strictly monotone\em: If $\mu$ measures $\{y\}$,
then $x\sqsubseteq y$ and $\mu x=\mu y$ implies $x=y.$
There are many important cases,
such as powerdomains and fractals\cite{martin:fractals},
where the applicability of measurement is greatly heightened
by the fact that $\ker\mu$ need not consist of \em all \em maximal
elements. However, in this paper, we are only interested in 
the case $\ker\mu=\max(D)$, so from here on we \em assume \em 
that this is part of the definition of measurement. 

\begin{example}\em  Canonical measurements.
\label{standardMeasurements}
\begin{enumerate}
\item[(i)] $(\IR,\mu)$ the interval domain with the length measurement
  $\mu[a,b]=b-a$. \item[(ii)] $([\nat\rightharpoonup\nat],\mu)$ the
  partial functions on the naturals with
\[\mu f=|\mbox{dom}(f)|\]
where $|\cdot|:\Pnat\rightarrow[0,\infty)^*$ is the measurement on the
algebraic lattice $\Pnat$ given by
        \[|x|=1-\sum_{n\in x}\frac{1}{2^{n+1}}.\]
      \item[(iii)] $(\Sigma^\infty,1/2^{|\cdot|})$ the Cantor set
        model where $|\cdot|:\Sigma^\infty\rightarrow[0,\infty]$ is
        the length of a string.         
\item[(iv)] $(\UX,\mbox{diam})$ the
        upper space of a locally compact metric space $(X,d)$ with
\[\mbox{diam}\,K=\sup\{d(x,y):x,y\in K\}.\]
\end{enumerate}
In each case, we have $\ker\mu=\max(D).$
\end{example}

We have previously seen how
order can implicitly capture
topology. With the addition of measurement,
we can also describe rates of change.
We restrict ourselves to
an extremely brief discussion of this.
 
\begin{definition}\em The $\mu$ \em topology \em on a continuous dcpo $D$ has as a basis all sets of the form
$\dua x\:\cap\downarrow\!\!y$, for $x,y\in D.$
\end{definition}
A sequence $(x_n)$ converges to $x$ in the $\mu$ topology iff it converges to $x$ in the Scott
topology and $(\exists n)\,x_k\sqsubseteq x,$ for all $k\geq n.$ In this case,
the largest tail of $(x_n)$ bounded by $x$ has $x$ as its supremum -- even
though $(x_n)$ may not be directed.

\begin{definition}\em Let $D$ be a domain with a map $\mu:D\rightarrow[0,\infty)^*$
that measures $X\subseteq D.$ If $f:D\rightarrow D$ is a partial map and
$p\in X\cap\dom(f)$ is not an isolated point of $\dom(f)$, then
\[df_\mu(p):=\lim_{x\rightarrow p}\frac{\mu f(x)-\mu f(p)}{\mu x-\mu p}\]
is called \em the informatic derivative \em of $f$ at $p$ with respect to $\mu$,
provided that it exists, as a limit in the $\mu$ topology.
\end{definition}
If the limit above exists, then it is unique, since the $\mu$ topology is Hausdorff,
and we are taking a limit at a point that is not isolated. Notice too the
importance of strict monotonicity of $\mu$: It ensures $\mu x-\mu p>0$. 
As with the upper space $\UX,$
a continuous $f:\reals\rightarrow\reals$ induces a Scott continuous map
\[\bar{f}:\IR\rightarrow\IR::x\mapsto f(x)\]
The following is proven in\cite{martin:thesis}.
\begin{theorem} If $f^\prime(p)$ exists, then $d\bar{f}_\mu[p]=|f^\prime(p)|.$
\end{theorem}
Interestingly, the informatic derivative on $\IR$ 
is \em equivalent \em to the
classical derivative for $C^1$ maps despite
the fact that it strictly generalizes it.

\section{Domains of classical and quantum states}

We now consider the domain of $n$ dimensional mixed states $\Omega^n$
in their spectral order. This order makes use of a simpler
domain of $n$ dimensional classical states $\Delta^n$ in their Bayesian order. After
introducing these domains, we show
how they can be used to provide order theoretic derivations
of the classical and quantum logics\cite{birkhoff}. Natural
measurements in these cases are the entropy functions
of Shannon and von Neumann. Thus, $\Delta^n$ and $\Omega^n$
fall right into line with
the examples of the last section. Despite this,
these domains are not continuous. They 
do possess a notion of approximation, though, 
which we discuss in the next section.

\subsection{Classical states} 

\begin{definition}\em Let $n\geq 2$. 
The \em classical states \em are 
\[\Delta^n:=\left\{x\in[0,1]^n:\sum_{i=1}^nx_i=1\right\}.\] 
A classical state $x\in\Delta^n$ is \em pure \em when $x_i=1$ for some 
$i\in\{1,\ldots,n\}$; we denote such a state by $e_i$. 
\end{definition} 

Pure states $\{e_i\}_i$ are the actual states a system can 
be in, while general mixed states $x$ and $y$ are epistemic entities. 
If we know $x$ and 
by some means determine that outcome $i$ is 
not possible, our knowledge 
improves to 
\[p_i(x)=\frac{1}{1-x_i}(x_1,\ldots,\hat{x_i},\ldots,x_{n+1})\in\Delta^n,\] 
where $p_i(x)$ is obtained 
by first removing $x_i$ from $x$ and 
then renormalizing. The partial mappings which result, 
\[p_i:\Delta^{n+1}\rightharpoonup\Delta^n\] 
with dom$(p_i)=\Delta^{n+1}\setminus\{e_i\}$, 
are called the {\it Bayesian projections\,} and lead 
one directly to the following relation on classical states. 

\begin{definition}\em For $x,y\in\Delta^{n+1}$, 
\begin{equation}\label{inductiverule} 
x\sqsubseteq y\equiv(\forall 
i)(x,y\in\mbox{dom}(p_i)\Rightarrow p_i(x)\sqsubseteq p_i(y)). 
\end{equation} 
For $x,y\in\Delta^2$, 
\begin{equation}\label{twostates} 
x\sqsubseteq y\equiv (y_1\leq x_1\leq 1/2)\mbox{ or }(1/2\leq x_1\leq
y_1)\,. 
\end{equation} 
The relation $\sqsubseteq$ on $\Delta^n$ is called the \em Bayesian 
order. \em 
\end{definition} 

To motivate (\ref{inductiverule}), if $x\sqsubseteq y$, 
then observer $x$ knows less than observer $y$. If 
something transpires which enables 
each observer to rule out exactly $e_i$ as a possible state of the system, 
then the first now knows $p_i(x)$, 
while the second knows $p_i(y)$. But since 
each observer's knowledge has increased by the 
same amount, the 
first must \em still \em know less than the second: 
$p_i(x)\sqsubseteq p_i(y).$ 

The order on two states (\ref{twostates}) is derived from 
the graph of Shannon entropy $\mu$ on $\Delta^2$ (left) as follows: 
\vspace{1.8mm} 
\begin{center} 
\begin{picture}(60,40) 
\qbezier(0,0)(20,60)(40,0) 
\put(0,0){\vector(1,0){50}} 
\put(0,0){\vector(0,1){40}} 
\put(-10,35){\normalsize$\mu$} 
\put(45,5){\normalsize$x_1$} 
\end{picture} 
$\stackrel{flip}{\longrightarrow}$ 
\begin{picture}(60,43) 
\qbezier(0,40)(20,-20)(40,40) 
\put(-10,45){\scriptsize$(1,0)$} 
\put(32,45){\scriptsize$(1,0)$} 
\put(17.2,0){\scriptsize$\bot=({1\over2},{1\over2})$} 
\end{picture} 
$\stackrel{pull}{\longrightarrow}$ 
\begin{picture}(40,43) 
\put(0,40){\line(3,-5){20}} 
\put(20,6.5){\line(3,5){20}} 
\put(-10,45){\scriptsize$(1,0)$} 
\put(32,45){\scriptsize$(1,0)$} 
\put(16.9,-1){\scriptsize$\bot=({1\over2},{1\over2})$} 
\end{picture} 
\end{center} 
\vspace{-1.3mm}\par\noindent 
The pictures above yield a canonical order on $\Delta^2$: 
\begin{theorem}\label{mixing} There 
is a unique partial order on $\Delta^2$ which has $\bot:=(1/2,1/2)$ and
satisfies the mixing law 
$$x\sqsubseteq y\ \mbox{and}\ p\in[0,1]\ \Rightarrow\ 
x\sqsubseteq(1-p)x+py\sqsubseteq y\,.$$ 
It is the Bayesian order on classical two states. 
\end{theorem} 

The \em least element \em in a poset is denoted 
$\bot$, when it exists. A more in depth derivation of the order is 
in\cite{meandbob}. 

\begin{theorem} $(\Delta^n,\sqsubseteq)$ is a dcpo with 
maximal elements 
\[\mathrm{max}(\Delta^n)=\{e_i:1\leq i\leq n\}\] 
and least element $\bot:=(1/n,\ldots,1/n)$. 
\end{theorem} 

The Bayesian order can also be described in a 
more direct manner, the \em symmetric characterization. \em 
Let $S(n)$ denote the group of permutations on $\{1,\ldots,n\}$ 
and \[{\Lambda^n:=\{x\in\Delta^n:(\forall i<n)\,x_i\geq x_{i+1}\}}\] denote
the collection of \em monotone \em 
classical states. 

\begin{theorem} 
\label{classicalsymmetries} 
For $x,y\in\Delta^n$, we have $x\sqsubseteq y$ iff 
there is a permutation ${\sigma\in S(n)}$ such that 
$x\cdot\sigma,y\cdot\sigma\in\Lambda^n$  
and 
\[(x\cdot\sigma)_i(y\cdot\sigma)_{i+1}\leq
(x\cdot\sigma)_{i+1}(y\cdot\sigma)_i\] 
for all $i$ with $1\leq i<n$. 
\end{theorem} 

Thus, the Bayesian order
is order isomorphic to $n!$ many copies of $\Lambda^n$
identified along their common boundaries. This fact,
together with the pictures of $\uparrow\!\!x$ and $\downarrow\!\!x$ at 
representative states $x$ in Figure 1, will
give the reader a good feel for the geometric
nature of the Bayesian order.

\begin{figure}[h] 
\centering\epsfig{figure=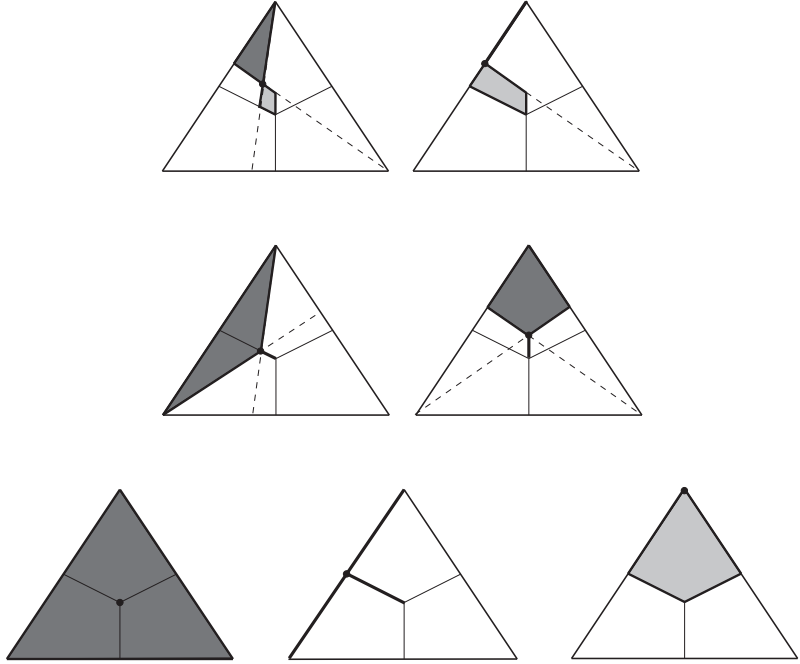,width=300pt} 
\caption{Pictures of $\uparrow\!\!x$ and $\downarrow\!\!x$ for $x\in\Delta^3$.} 
\end{figure} 

\subsection{Quantum states} 

Let $\mathcal{H}^n$ denote an $n$-dimensional complex Hilbert 
space with specified inner product $\langle\cdot|\cdot\rangle.$ 

\begin{definition}\em A \em quantum state \em is 
a density operator ${\rho:\mathcal{H}^n\rightarrow\mathcal{H}^n}$, i.e., 
a self-adjoint, positive, linear operator with $\mathrm{tr}(\rho)=1.$ 
The quantum states on $\mathcal{H}^n$ are denoted $\Omega^n$. 
\end{definition} 

\begin{definition}\em A quantum state $\rho$ on $\mathcal{H}^n$ 
is \em pure \em if 
\[\mathrm{spec}(\rho)\subseteq\{0,1\}.\] 
The set of pure states is denoted $\Sigma^n$. They 
are in bijective correspondence with the one dimensional subspaces 
of $\mathcal{H}^n.$ 
\end{definition} 

Classical states are distributions on the 
set of pure states $\max(\Delta^n).$ 
By Gleason's theorem\cite{Gleason}, an analogous 
result holds for quantum states: Density operators encode 
distributions on the set of pure states $\Sigma^n$ 
up to equivalent behavior under measurements. 

\begin{definition}\em A \em quantum observable \em is 
a self-adjoint linear operator ${e:\mathcal{H}^n\rightarrow\mathcal{H}^n}.$ 
\end{definition} 

If our 
knowledge about the state of the system 
is represented by density operator $\rho$, 
then quantum mechanics predicts 
the probability that a measurement of observable $e$ 
yields the value $\lambda\in\mathrm{spec}(e)$. It is 
\[\mathrm{pr}(\rho\rightarrow
e_\lambda):=\mathrm{tr}(p_e^\lambda\cdot\rho),\] 
where $p_e^\lambda$ is the projection corresponding 
to eigenvalue $\lambda$ and $e_\lambda$ is its associated 
eigenspace in the \em spectral representation \em of $e$. 

\begin{definition}\em Let $e$ be an observable on $\mathcal{H}^n$ with 
$\mathrm{spec}(e)=\{1,\ldots,n\}.$ For a quantum state $\rho$ on 
$\Omega^n$, 
\[\mathrm{spec}(\rho|e):=(\mathrm{pr}(\rho\rightarrow
e_1),\ldots,\mathrm{pr}(\rho\rightarrow 
e_n))\in\Delta^n.\] 
\end{definition} 

For the rest of the paper, we 
assume that all observables $e$ have $\mathrm{spec}(e)=\{1,\ldots,n\}.$ 
For our purposes it is enough to assume $|\mathrm{spec}(e)|=n$; 
the set $\{1,\ldots,n\}$ is chosen for the sake of aesthetics. 
Intuitively, then, $e$ is an experiment 
on a system which yields 
one of $n$ different outcomes; if our a priori knowledge about the state of
the system 
is $\rho$, then our knowledge about 
what the result of experiment $e$ \em will be \em is 
$\mathrm{spec}(\rho|e).$ Thus, $\mathrm{spec}(\rho|e)$ determines 
our ability to \em predict \em the result 
of the experiment $e$. 

So what does it mean to say that we have more information 
about the system when we have $\sigma\in\Omega^n$ than when we have
$\rho\in\Omega^n$? 
It could mean that there is an experiment 
$e$ which (a) serves as a physical realization of the knowledge 
each state imparts to us, and (b) that we have a better 
chance of predicting the result of $e$ from state $\sigma$ than we do from
state $\rho$. 
Formally, (a) means 
that $\mathrm{spec}(\rho)=\mathrm{Im}(\mathrm{spec}(\rho|e))$ and 
$\mathrm{spec}(\sigma)=\mathrm{Im}(\mathrm{spec}(\sigma|e))$, 
which is equivalent to 
requiring $[\rho,e]=0$ and $[\sigma,e]=0$, 
where $[a,b]=ab-ba$ is the commutator of operators. 

\begin{definition}\em Let $n\geq 2$. For quantum states 
$\rho,\sigma\in\Omega^n$, we have $\rho\sqsubseteq \sigma$ iff there 
is an observable $e:\mathcal{H}^n\rightarrow\mathcal{H}^n$ such 
that $[\rho,e]=[\sigma,e]=0$ and 
$\mathrm{spec}(\rho|e)\sqsubseteq\mathrm{spec}(\sigma|e)$ in $\Delta^n$. 
\end{definition} 

This is called the \em spectral order \em on quantum states. 

\begin{theorem} $(\Omega^n,\sqsubseteq)$ is a dcpo with maximal elements 
\[\mathrm{max}(\Omega^n)=\Sigma^n\] 
and least element $\bot=I/n$, where $I$ is 
the identity matrix. 
\end{theorem} 

There is one case where the spectral order 
can be described in an elementary manner. 

\begin{example}\em As is well-known, the $2\times 2$ density operators can 
be represented as points on the unit ball in $\reals^3:$ 
\[\Omega^2\simeq\{(x,y,z)\in\reals^3:x^2+y^2+z^2\leq 1\}.\] 
For example, the origin $(0,0,0)$ corresponds 
to the completely mixed state $I/2$, 
while the points on the surface of the sphere 
describe the pure states. The order on $\Omega^2$ then amounts 
to the following: $x\sqsubseteq y$ iff the line 
from the origin $\bot$ to $y$ passes through $x$. 
\end{example} 

Like the Bayesian order on $\Delta^n$, 
the spectral order on $\Omega^n$ can also 
be characterized in terms 
of symmetries and projections. In its symmetric 
formulation, \em unitary operators \em on $\mathcal{H}^n$ 
take the place of permutations on $\{1,\ldots,n\}$, 
while the projective formulation of $(\Omega^n,\sqsubseteq)$ 
shows that each classical projection
$p_i:\Delta^{n+1}\rightharpoonup\Delta^n$ 
is actually the restriction of a special 
quantum `projection' $\Omega^{n+1}\rightharpoonup\Omega^k$ 
with $k=n$. 

\subsection{The logics of Birkhoff and von Neumann} 

The logics of Birkhoff and von Neumann\cite{birkhoff}
consist of the propositions one can make about 
a physical system. Each proposition takes the form 
``The value of observable $e$ is contained in 
$E\subseteq\mathrm{spec}(e).$'' For 
classical systems, the logic 
is $\mathcal{P}\{1,\ldots,n\}$, 
while for quantum systems it is $\mathbb{L}^n$, 
the lattice of (closed) subspaces of $\mathcal{H}^n.$ 
In each case, implication of propositions is 
captured by inclusion, and a fundamental 
distinction between 
classical and quantum -- that 
there are pairs of quantum observables 
whose exact values cannot be simultaneously measured at a single 
moment in time -- finds lattice theoretic expression:
$\mathcal{P}\{1,\ldots,n\}$ is distributive; 
$\mathbb{L}^n$ is not. 

We now 
establish the relevance of the 
domains $\Delta^n$ and $\Omega^n$ 
to theoretical physics: The classical 
and quantum logics can be \em derived \em 
from the Bayesian and spectral orders 
using the \em same \em order theoretic technique. 

\begin{definition}\em An element $x$ of a dcpo $D$ is \em 
irreducible \em when 
\[\bigwedge(\uparrow\!\!x\cap\max(D))=x\] 
The set of irreducible elements in $D$ is written $\mbox{Ir}(D).$ 
\end{definition} 

The order dual of a poset $(D,\sqsubseteq_D)$ is written $D^*$; its 
order is $x\sqsubseteq y\Leftrightarrow y\sqsubseteq_D x.$ 

\begin{theorem} For $n\geq 2$, the classical lattices arise as 
\[\mathrm{Ir}(\Delta^n)^*\simeq\mathcal{P}\{1,\ldots,n\}\setminus\{\emptyset\},\] 
and the quantum lattices arise as 
\[\mathrm{Ir}(\Omega^n)^*\simeq\mathbb{L}^n\setminus\{0\}.\] 
\end{theorem} 

It is worth pointing out 
that these logics consist exactly of 
the states traced out by
the motion of a searching process on
each of the respective
domains. To illustrate, 
let ${p_i^+:\Delta^n\rightarrow\Delta^{n}}$ for $1\leq i\leq n$ 
denote the result of first applying the Bayesian projection $p_i$ to a state,
and then reinserting a zero in place of the element removed.
Now, beginning with $\bot\in\Delta^n$, apply
one of the $p_i^+$. This projects 
away a single outcome from $\bot$, leaving us
with a new state. For the
new state obtained, project away another 
single outcome; after $n-1$ iterations, this process 
terminates with a pure state $e_i$, and all 
the intermediate states comprise a path from $\bot$ to $e_i$. 
Now imagine all the possible paths from $\bot$ to a pure state which
arise in this manner. This set of states is exactly $\mathrm{Ir}(\Delta^n).$ (See Figure 2). 

\begin{figure}[h] 
\centering\epsfig{figure=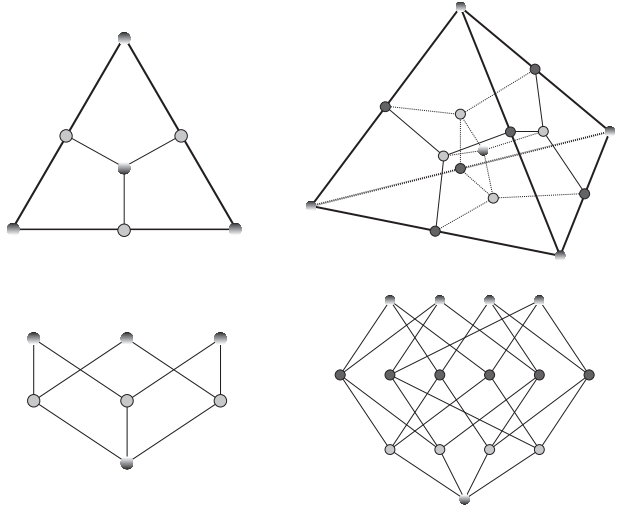,width=230pt} 
\caption{The irreducibles of $\Delta^3$ and $\Delta^4$ with their 
corresponding Hasse diagrams.} 
\end{figure} 


\subsection{Entropy} 

The formal notion of information content studied in measurement
is broad enough in scope to capture Shannon's idea from information
theory, as well as von Neumann's conception of entropy from quantum mechanics.

\begin{theorem}Shannon entropy 
\[\mu x=-\sum_{i=1}^nx_i\log x_i\] 
is a 
measurement of type $\,\Delta^n\rightarrow[0,\infty)^*.$ 
\end{theorem} 

A more subtle example of a measurement on classical states 
is the retraction $r:\Delta^n\rightarrow\Lambda^n$ which 
rearranges the probabilities in a classical state into 
descending order.

\begin{theorem} von Neumann entropy 
\[\sigma\rho = -\mathrm{tr}(\rho\log\rho)\] 
is a measurement of type $\Omega^n\rightarrow[0,\infty)^*.$ 
\end{theorem} 

Another natural measurement on $\Omega^n$ is the map 
${q:\Omega^n\rightarrow\Lambda^n}$ which assigns to 
a quantum state its spectrum rearranged into descending order. 
It is an important
link between classical 
and quantum information theory. 

By combining the quantitative and qualitative aspects 
of information, we obtain a highly effective method
for solving a wide range of problems in the sciences. As an 
example, consider the problem of \em rigorously \em proving
the statement ``there is more information in the quantum than in
the classical.'' 

The first step is to think carefully
about why we say
that the classical is contained in the quantum; one
reason is that for any observable $e$, 
we have an isomorphism 
\[\Omega^n|e=\{\rho\in\Omega^n:[\rho,e]=0\}\simeq\Delta^n\]
between the spectral and Bayesian orders. That is, each classical state can be assigned to a quantum state
in such a way that \em information is conserved\em:
\begin{center}
\mbox{conservation of information}\\
$=$\\
\mbox{\ (qualitative conservation)} + \mbox{(quantitative
conservation)}\\
$=$\\
\mbox{\ \ \ \ \ \ \ \ \ \ (order embedding)} + \mbox{(preservation of entropy)}.
\end{center}
This realization, that both the qualitative
\em and \em the quantitative characteristics of
information are preserved in passing from the classical
to the quantum, solves the problem.

\begin{theorem} Let $n\geq 2$. Then 
\begin{enumerate} 
\em\item[(i)]\em There is an order embedding 
$\phi:\Delta^n\rightarrow\Omega^n$ with ${\sigma\circ\phi=\mu.}$ 
\em\item[(ii)]\em For any $m\geq 2$, there is no order 
embedding ${\phi:\Omega^n\rightarrow\Delta^m}$ with
${\mu\circ\phi=\sigma.}$ 
\end{enumerate} 
\end{theorem} 
Part (ii) is true for any pair of
measurements $\mu$ and $\sigma$. The proof is fun: If (ii) is false,
then $\phi$ restricts to an injection of $\max(\Omega^n)$ into $\max(\Delta^n)$,
using $\ker\mu\subseteq\max(\Delta^n)$ and
$\ker\sigma=\max(\Omega^n)$. But no such injection
can actually exist: $\max(\Omega^n)$ is infinite, $\max(\Delta^n)$ is not.

\section{Axioms for partiality}

We have already mentioned that the domains
$(\Delta^n,\sqsubseteq)$ and $(\Omega^n,\sqsubseteq)$ are not
continuous. The easiest way to see why is
to take note of the fact that the Bayesian order on $\Delta^n$
is \em degenerative\em: If $x\sqsubseteq y$, then
\[y_i=y_j>0 \Rightarrow x_i=x_j>0.\]
Using this property, it is easy to show that the only
approximation of a
state like $(1/2,1/2,0)$ is $\bot$
by
construct an increasing
sequence $(y_n)$ 
whose last two components are equal
such that ${(1/2,1/2,0)\sqsubseteq e_1=\bigsqcup y_n}$.
Nevertheless, these
domains do possess a notion of approximation.

 \begin{definition}\em Let $D$ be a dcpo. For $x,y\in D$, we
write $x\ll y$ iff for all directed sets $S\subseteq D$,
\[y=\bigsqcup S\Rightarrow (\exists s\in S)\,x\sqsubseteq s.\]
The \em approximations \em of $x\in D$ are
\[\dda x:=\{y\in D:y\ll x\},\]
and $D$ is called \em exact \em if $\dda x$
is directed with supremum $x$ for all $x\in D$.
\end{definition}

Notice that the difference between this definition and
the previous is that $\sqsubseteq$ has been replaced with `$=$'.
A \em continuous dcpo \em is exact, for example, 
and in that
case, the classical definition of $\ll$ is equivalent
to the one above. The following is proven in\cite{meandbob}:

\begin{theorem} 
$(\Delta^n,\sqsubseteq)$ and $(\Omega^n,\sqsubseteq)$ are exact.
\end{theorem}
To hint at why, we
can approximate any $x\in\Delta^n$ using
the \em straight line path \em $\pi_{\bot x}:[0,1]\rightarrow\Delta^n$ 
from $\bot$ to $x$,
\[\pi_{\bot x}(t)=(1-t)\bot+tx.\]
It is Scott continuous with $\pi_{\bot x}(t)\ll x$ for $t<1$. The
analogous result holds for $\Omega^n.$

\begin{definition}\em An element $x\in D$ is a \em coordinate \em if 
either $x\in\mathrm{Ir}(D)$ or $x\in\dda\mathrm{Ir}(D).$
\end{definition}
In the case of $\Delta^n$ and $\Omega^n$, a coordinate
is either a \em proposition \em or an \em approximation of a proposition. \em
Equivalently, a coordinate is a state
on one of the lines joining $\bot$ to a proposition.  

\begin{theorem}
Each state is the supremum of coordinates.
\end{theorem}  

The result above, proven in\cite{coecke},
holds for both $\Delta^n$ and $\Omega^n$. We do not
expect all domains to have this property,
but the role of partiality in defining
`coordinate' -- as either an irreducible
or an \em approximation \em of an irreducible
-- may be worth taking note of in trying to
develop a general and useful
set of axioms for the description of partiality.
Ideally, these axioms will
\begin{itemize}  
\item generalize continuous domains,
\item include $(\Delta^n,\sqsubseteq)$ and $(\Omega^n,\sqsubseteq)$
as examples, 
\item aid in the description of a fundamental topology,
which will be equivalent to the Scott topology in the case of continuous dcpo's,
and
\item be relatable to implicit
uses of the notion in physics,
such as `dynamics' (i.e., causality relations on light cones\cite{bomb}).
\end{itemize}
The interested reader will notice
that exact dcpo's definitely satisfy
the first two criteria. We do not know about
the other two (or even what the last one
may mean). Nevertheless, we hope
this paper will serve as a useful
guide for those intent on looking.


\end{document}